\documentclass[prb,a4paper,eqsecnum,preprint]{revtex4}
\usepackage[pdftex]{hyperref}
\usepackage[fleqn]{amsmath}
\usepackage{amsfonts}
\usepackage{amssymb}
\usepackage{graphicx}

\newcommand{\todo}[1]{\marginpar{#1}}

\usepackage[OT2,OT1]{fontenc}
 \newcommand\cyr{%
 \renewcommand\rmdefault{wncyr}%
 \renewcommand\sfdefault{wncyss}%
 \renewcommand\encodingdefault{OT2}%
 \normalfont
 \selectfont}
 \DeclareTextFontCommand{\textcyr}{\cyr} 
\newcommand{\cha}{\textcyr{W}}

\newcommand{\ie}{{\em i.e.\/}} 
\newcommand{\eg}{{\em e.g.\/}}

\def \dd{\hbox{d}}

\newcommand{\vect}[1]{\boldsymbol{#1}}

\newcommand{\vm}{\vect{m}}

\newcommand{\M}{_{\text{\footnotesize M}}}
\newcommand{\vn}{\vect{n}}
\newcommand{\ZZ}{\mathbb{Z}}
\newcommand{\MM}{\mathbb{M}}

\newcommand{\ds}{\displaystyle}
\newcommand{\TF}[1]{\widetilde{#1}}

\newcommand{\bb}{\begin{equation}}
\newcommand{\ee}{\end{equation}}

\newcommand{\ver}{\vect{r}}

\newcommand{\vk}{\vect{k}}

\newcommand{\TFM}{\TF{\mathbb{M}}}
\newcommand{\KK}{\mathbb{K}}
\newcommand{\PPPM}{\mathrm{P3M}}

\newcommand{\moy}[1]{\left\langle{#1}\right\rangle}

\begin{document}
\title{The optimal P3M algorithm for computing
  electrostatic energies in periodic systems}

\author{V. Ballenegger} 
\affiliation{Institut UTINAM, Universit\'e de Franche-Comt\'e, UMR
  6213, 16, route de Gray, 25030 Besan\c{c}on cedex \textsc{France}.}\
\author{J. J. Cerda} 
\affiliation{Frankfurt Inst. for Advanced Studies, J.W. Goethe -
  Universit\"{a}t, Frankfurt, Germany}
\author{Ch. Holm} 
\affiliation{Frankfurt Inst. for Advanced Studies, J.W. Goethe -
  Universit\"{a}t, Frankfurt, Germany}
\affiliation{Max-Planck-Institut f\"{u}r Polymerforschung, Mainz,
  Germany}
\author{O. Lenz}
\affiliation{Frankfurt Inst. for Advanced Studies, J.W. Goethe -
  Universit\"{a}t, Frankfurt, Germany}

\date{\today}

\begin{abstract}
  We optimize Hockney and Eastwood's Particle-Particle Particle-Mesh
  (P3M) algorithm to achieve maximal accuracy in the electrostatic
  energies (instead of forces) in 3D periodic charged systems. To this
  end we construct an optimal influence function that minimizes the
  RMS errors in the energies. As a by-product we derive a new
  real-space cut-off correction term, give a transparent derivation of
  the systematic errors in terms of Madelung energies, and provide an
  accurate analytical estimate for the RMS error of the energies. This
  error estimate is a useful indicator of the accuracy of the computed
  energies, and allows an easy and precise determination of the
  optimal values of the various parameters in the algorithm (Ewald
  splitting parameter, mesh size and charge assignment order).
\end{abstract}  

\maketitle   

\section{Introduction}
Long range interactions are ubiquitously present in our daily life.
The calculation of these interactions is, however, not an easy task to
perform.  One needs indeed to resort to specialized algorithms to
overcome the quadratic scaling with the number of particles, as soon
as the simulated system includes more than a few hundred particles,
see for example the review of Arnold and Holm \cite{arnold05a}.  In Molecular
Dynamics simulations, one is mainly interested in the accuracy of the
force computation, since they govern the dynamics of the system.  In
contrast, in Monte Carlo (MC) simulations, the concern is to compute
accurate energies.  If the potential is of long range (\eg{} a Coulomb
potential or dipolar interaction), and one has chosen to use periodic
boundary conditions, the computation of both observables is quite time
consuming if one uses the traditional Ewald sum. Since the seminal
work of Hockney and Eastwood~\cite{HE} it has been common to resort to
a faster way of calculating the reciprocal space sum in the Ewald
method with the help of Fast-Fourier-Transforms (FFTs). These
algorithms are called mesh-based Ewald sums, and various variants
exist \cite{DH}. They all scale as $N\log N$ with the number of
charged particles $N$, and the algorithms are nowadays routinely used
in simulations of bio-systems, charged soft matter, plasmas, and many
more areas.  The most accurate variant is still the original method of
Hockney and Eastwood, which they called
particle-particle-particle-mesh (P3M), and into which various other
improvements like the analytical differentiation used in other
variants of the mesh-based Ewald sum~\cite{SPME} can be built in. In
addition, an accurate error estimate for P3M exists, so that one can
tune the algorithm to a preset accuracy, thus maximizing the
computational efficiency before doing any simulations \cite{DH2}.

While in the standard P3M algorithm\cite{HE}, the lattice Green
function, called the ``influence function'', is optimized to give the
best possible accuracy in the forces, the electrostatic energy is
usually calculated with the same force-optimized influence
function. However, there are certainly situations where one needs a
high precision of the energies, for instance in Monte Carlo simulations, and
the natural question arises whether one can optimize the influence
function to enhance the accuracy of the P3M energies. The main goal of
this paper is to derive the energy-optimized influence function, and
to derive an analytical estimate for the error in the P3M energies.
This error estimate is a valuable indicator of the accuracy of the
calculations and allows a straightforward and precise determination of
the optimal values of the various parameters in the algorithm (Ewald
splitting parameter, mesh size, charge assignment order).

The present derivation of the optimal influence function, and the
associated error estimate, is concise and entirely self-contained. The
present paper can thus also serve as a pedagogical introduction to the
main ideas and mathematics of the P3M algorithm.

The paper is organized as follows. In Sec.~\ref{Ewald}, we briefly
review the ideas of the standard Ewald method and provide the most
important formulae.  In Sec.~\ref{Ewald_correction}, we derive direct
and reciprocal space correction terms which compensate, on average,
the effects of cut-off errors in the standard Ewald method. We
interpret the formulae in terms of the direct and reciprocal space
components of the Madelung energies of the ions. In
Sec.~\ref{mesh}, the calculation of the reciprocal energy
according to the P3M algorithm (\ie{} with a fast Fourier transform and
an optimized influence function) is presented.  The mathematical
analysis of the errors introduced by the discretization on a grid is
performed in Sec.~\ref{mesh_errors}. This analysis is used in
Sec.~\ref{p3m_optimization} to derive the energy-optimized influence
function and the associated RMS error estimate. The derivation shows
that the P3M energies must be shifted to compensate for systematic
cut-off and aliasing errors in the Madelung energies of the ions.
Finally, our analytical results are tested numerically in
Sec.~\ref{p3m_check}.

\section{The Ewald sum} 
\label{Ewald}

We consider a system of $N$ particles with charges $q_i$ at positions
$\vect{r}_i$ in an overall neutral and (for simplicity) cubic
simulation box of length~$L$ and volume $V=L^3$. If periodic boundary
conditions are applied, the total electrostatic energy of the box is
given by
\begin{equation} \label{Ewald_sum} 
E = \frac{1}{2}\sum_{\vn\in\ZZ^3} \sideset{}{^{'}}\sum_{i,j=1}^N
q_i q_j v(\ver_{ij}+\vn L), 
\end{equation} 
where $v(\ver) = {1}/{|\ver|}$ is the Coulomb potential,
$\ver_{ij}=\ver_i-\ver_j$, and $\vn$ is a vector with integer
components that indexes the periodic images. The prime indicates that
the (divergent) summand for $i=j$ has to be omitted when $\vn=0$.

Because of the slow decay of the Coulomb interaction, the sum
in~\eqref{Ewald_sum} is only conditionally convergent: its value is
not well defined unless one specifies the precise way in which the
cluster of simulation boxes is supposed to fill~$\mathbb{R}^3$. Often,
one chooses a spherical order of summation, which is equivalent to the
limit of a large, spherically bounded, regular grid of replicas of the
simulation box, embedded in vacuum. The simulation box can then be
pictured as the central \textsc{LEGO} brick in a huge ball made up of
such bricks. If this ``lego ball'' is surrounded by a homogeneous
medium with dielectic constant~$\epsilon'$ ($\epsilon' = 1$ if it's
vacuum) and if the simulation box has a net dipole moment
$\vect{M}=\sum_i q_i \ver_i$, the particles in the ball will feel a
depolarizing field created by charges that appear on the surface of
the uniformily polarized ball. It can be shown that the work done
against this depolarizing field when charging up the system is
\begin{equation}	\label{E^d}
   E^{(d)} = \frac{2\pi \vect{M}^2}{(1+2\epsilon')L^3}
\end{equation}
in the case of a spherical order of summation~\cite{deLeeuw,Caillol}
(for other summation orders, see the articles of Smith\cite{Smith88}
and Ballenegger and Hansen \cite{BalHan}). The energy~$E^{(d)}$ is
contained, even if not easily seen, in the total electrostatic
energy~\eqref{Ewald_sum} (at least when $\epsilon'=1$ since such a
vacuum boundary condition was assumed in writing~\eqref{Ewald_sum}).
Obviously, the energy $E^{(d)}$ vanishes if we employ metallic
boundary conditions defined by $\epsilon'=\infty$.
  
The fact that $E^{(d)}$ depends on the order of summation, and hence
on the shape of macroscopic sample under consideration, is a
consequence of the conditional convergence of the
sum~\eqref{Ewald_sum}. Due to the energy cost~$E^{(d)}$, the
fluctuations of the total dipole moment of the simulation box (and
hence of the considered macroscopic sample) depend on the dielectric
constant~$\epsilon'$ and on the shape of the sample. The
energy~$E^{(d)}$ is crucial to ensure, for example, that the
dielectric constant~$\epsilon$ of the simulated system obtained from
the Kirkwood formula~\cite{Kirkwood}, which relates $\epsilon$ to the
fluctuations of the total dipole moment, is independent of the choices
made for the sample shape and for the dielectric boundary
condition~\cite{AlaBal,BalHan2}.


Ewald's method to compute the energy~\eqref{Ewald_sum} is based on a
decomposition of the Coulomb potential,
$v(\ver)=\psi(\ver)+\phi(\ver)$, such that $\psi(\ver)$ contains the
short-distance behavior of the interaction, while $\phi(\ver)$
contains the long-distance part of the interaction and is regular at
the origin. The traditional way to perform this splitting is to
define  
\begin{equation} \label{def_phi} 
  \phi(\ver) = \mathrm{erf}(\alpha r)/r, \qquad r = |\ver|,
\end{equation} 
and 
\begin{equation} \label{def_psi}
  \psi(\ver)= v(\ver)-\phi(\ver) = \mathrm{erfc}(\alpha r)/r
\end{equation}
With this choice, $\psi(r)$ corresponds to the interaction energy
between a unit charge at a distance~$r$ from another unit charge that
is screened by a neutralizing Gaussian charge distribution whose width
is controlled by the Ewald length~$\alpha^{-1}$. Following this
decomposition of the potential, the electrostatic energy can be
written in the well-known Ewald form~\cite{Ewald21,deLeeuw}: 
\begin{equation}	\label{Ewald_form}
  E = E^{(r)} + E^{(k)} + E^{(d)}
\end{equation}
where the {\it real-space energy} $E^{(r)}$ contains the contributions
from short-range interactions~$\psi(\ver)$, \ie
\begin{equation}
\label{U^r} E^{(r)} = \frac{1}{2}\sum_{\vn\in\ZZ^3}\sideset{}{^{'}}\sum_{i,j=1}^N
q_i q_j \psi(\ver_{ij}+\vn L),
\end{equation}
and the {\it reciprocal space energy} $E^{(k)}$ contains contributions
from long-range interactions~$\phi(\ver)$ (apart from the
contributions that are responsible for the conditional convergence
which are included in the term~$E^{(d)}$ in~\eqref{Ewald_form}). The
fact that the surface term (or ``dipole term'') $E^{(d)}$ is
independent of the Ewald parameter~$\alpha$ shows that this
contribution is not specific to the Ewald method, but more generally
reflects the problems inherent to the conditional convergence of the
$\vn$ sum in Eq.~\eqref{Ewald_sum}. Contrary to $E^{(r)}$ which can be
computed easily in real space thanks to the rapid decay of the~$\psi$
interaction, $E^{(k)}$ is best computed in Fourier space, where it can
be expressed as~\cite{Ewald21}
\begin{equation}	\label{2.8}
  E^{(k)} = E^{(ks)} - E^{(s)}
\end{equation}
where
\begin{align}
\label{E^k} E^{(ks)} &=
\frac{1}{2L^3}\sum_{\substack{\vk\in\KK\\\vk\neq0}}|\TF{\rho}(\vk)|^2
\TF{\phi}(\vk)  \\ \label{E^s}
E^{(s)} &= Q^2 \frac{\alpha}{\sqrt{\pi}}
\end{align}
with
\begin{equation} \label{def_Q^2}
Q^2 = \sum_{i=1}^N q_i^2.
\end{equation}
In~\eqref{E^k}, $\TF{\phi}(\vk)$ is the Fourier transform of
the reciprocal interaction~\eqref{def_phi}, 
\begin{equation} 
\label{TF_phi}
\TF{\phi}(\vk)=\int e^{-i\vk\cdot\ver} \phi(\ver)\dd\ver =
\frac{4\pi}{k^2}\exp(-k^2/4\alpha^2), 
\end{equation} 
and $\TF{\rho}(\vk)$ is the
Fourier transformed charge density 
\begin{equation} 
\label{def_TF_rho}
\TF{\rho}(\vk) = \sum_{i=1}^N q_i\, e^{-i\vk\cdot\ver_i}.  
\end{equation}
The sum in~\eqref{E^k} is over wave vectors in the discrete set
$\mathbb{K}=\{2\pi \vn/L:\,\vn\in\ZZ^3\}$. The term $\vk=0$ is
excluded in the sum because of the overall charge neutrality. The
self-energy term $E^{(s)}$ compensates for the self-energies (the
reciprocal interaction of each particle with itself $\frac{1}{2}q_i^2
\phi(\ver=\vect{0})=q_i^2 \alpha/\sqrt{\pi}$) that are included in
$E^{(ks)}$.

The energy~\eqref{Ewald_sum} converges only for
systems that are globally neutral. For systems with a net charge, the
sum can be made convergent by adding a homogeneously distributed
background charge which restores neutrality. In that case, an
additional contribution~\cite{Hummer}
\begin{equation} \label{E^n}
E^{(n)} = -\frac{\pi}{2 \alpha^2 L^3} \Big(\sum_{i=1}^N q_i\Big)^2
\end{equation}
must be added to~\eqref{Ewald_form} to account for the interaction
energies of the charges with the neutralizing background.

The reciprocal energy~$E^{(ks)}$, defined by the Ewald
formula~\eqref{2.8}, is the starting point of mesh-based Ewald sums,
which are methods to compute efficiently that energy in many-particle
systems. Notice that~\eqref{2.8} can also be written in an alternative
form in terms of a pair potential and a Madelung self-energy, see
Appendix~\ref{AA}. The inverse length $\alpha$ tunes the relative
weight of the real space $E^{(r)}$ and the reciprocal space $E^{(k)}$
contributions to the energy, but the final result is independent
of~$\alpha$. In practice, $E^{(r)}$ and $E^{(k)}$ can be computed
using cut-offs, because the sum over $\vect{n}$ in~\eqref{U^r} and the
sum over $\vect{k}$ in~\eqref{E^k} converge exponentially fast.
Typically, one chooses $\alpha$ large enough to employ the minimum
image convention\cite{Allen} in Eq.~\eqref{U^r}.

At given real and reciprocal space cut-offs $r_\mathrm{cut}$ and $k_{\rm
  cut}$, there exists actually an \emph{optimal}~$\alpha$ such that the
accuracy of the approximated Ewald sum is as high as possible. This
optimal value can be determined with the help of the estimates for the
cut-off errors derived by Kolafa and Perram~\cite{KolPer}, by demanding
that the real and reciprocal space contributions to the error are
equal.  Kolafa and Perram's root-mean-square error estimates are
\begin{equation} 
\label{RMS_U^r} 
\Delta E^{(r)} \simeq Q^2 \sqrt{\frac{r_c}{2L^3}}
\frac{e^{-\alpha^2 r^2_\mathrm{cut}}}{(\alpha r_\mathrm{cut})^2} 
\end{equation} 
and
\begin{equation} 
\label{Kolafa_DeltaU^k} 
\Delta E^{(k)} \simeq Q^2 \alpha
\,\frac{e^{-(\pi k_\mathrm{cut}/\alpha L)^2}}{\pi^2 k_\mathrm{cut}^{3/2}}.
\end{equation} 
These error estimates make explicit the exponential dependence of the
error on the real and reciprocal space cut-offs.

Formula~\eqref{Kolafa_DeltaU^k} is actually valid only when a
correction term (given by Eq.~\eqref{U^k_cut} below) is added, to
compensate the systematic error that affects the reciprocal energies
when the sum over wave vectors in~\eqref{E^k} is truncated. The origin
of this correction term is explained in detail in
Sec.~\ref{Ewald_correction}.  A similar term must also be introduced
in the P3M algorithm when one computes the electrostatic energy.
Similarly, the direct-space energy~\eqref{U^r} also contains a
systematic error when the pair-wise interaction is truncated at the
cut-off distance $r_\mathrm{cut}$. The derivation in the next section
will also provide a correction term for this effect.

To summarize, the final Ewald formula for the total electrostatic energy reads
\begin{equation}
  \begin{split}
    E & = E^{(r)} & \mathrm{[eq. \eqref{U^r}]}\\
    & + E^{(ks)} & \mathrm{[eq. \eqref{E^k}]}\\
    & - E^{(s)} & \mathrm{[eq. \eqref{E^s}]}\\
    & + E^{(d)} & \mathrm{[eq. \eqref{E^d}]}\\
    & + E^{(n)} & \mathrm{[eq. \eqref{E^n}]}\\
  \end{split}
\end{equation}
Furthermore, when the sums in $E^{(r)}$ and $E^{(k)}$ are evaluated
numerically using cut-offs, an additional correction term
$E_\mathrm{cut}$, defined in Eq.~\eqref{U_cut} below, must be added to
the truncated energy, as shown in the next section.

\section{Correction term for truncated Ewald sums} 
\label{Ewald_correction}

If we consider electroneutral systems where the charged particles are
located at random, we expect the electrostatic energy to vanish on
average, because there is an equal probability to find a positive or
negative charge at any relative distance~$r$. However, when periodic
boundary conditions (PBC) are applied, the average energy of random
systems does not vanish, because each charge interacts with its own
periodic images (and with the uniform neutralizing background provided by
the other charges). 

Since this interaction energy $E_\mathrm{img}$ of an ion with its
periodic images and with the neutralizing background does not depend on the position of the ion in the
simulation box, it plays the role of a ``self-energy''. We will refer
to $E_\mathrm{img}$ as the Madelung (self-)energy of an ion, to avoid
confusion with the self-energy $\frac{1}{2}q^2\phi(0)$ already defined
in the Ewald method as the reciprocal interaction of a particle with
itself. 

We denote by angular brackets the average over the positions of the
$N$ charged particles:
\begin{equation} 
\moy{\cdots} = \frac{1}{V^N} \int_{V^N} \cdots \;
\dd\ver_1 ...\, \dd\ver_N.  
\end{equation} 

\subsection{Madelung energy}

The Madelung energy of an ion takes the form
$E_\mathrm{img}=\frac{1}{2}q^2 \zeta$, where $\zeta$ is a purely
numerical factor in units of $1/L$ that depends only on the size and
shape of the simulation box.

Let us calculate the average electrostatic energy of random charged
systems in PBC, to find the value of $\zeta$ and derive a correction
term for cut-off errors in truncated Ewald sums (some results derived
here will be used in Sec.~\ref{S<DeltaU>}). 
On the one hand, the average Coulomb energy of the random systems is
by definition $Q^2\zeta/2$, while on the other hand, it can be
calculated as the sum of a direct space contribution $\moy{E^{(r)}}$
and a reciprocal space contribution $\moy{E^{(k)}}$. The average
reciprocal energy is, using~\eqref{E^s} and \eqref{E^k},
\begin{equation} 
\label{Calc_zeta^k} 
\moy{E^{(k)}} =
\frac{1}{2L^3}\sum_{i,j}q_i q_j\sum_{\substack{\vk\in\KK\\\vk\neq0}}
\moy{e^{-i\vk\cdot(\ver_i-\ver_j)}} \TF{\phi}(\vk) - Q^2
\frac{\alpha}{\sqrt{\pi}}.  
\end{equation} 
Since $\moy{\exp(-i\vk\cdot\ver_j)} = \delta_{\vk,\vect{0}}$, all
terms with $j \neq i$ vanish (this is due to the fact that the Ewald
pair potential averages to zero, see Appendix~\ref{AA}). By contrast,
``self'' terms ($i=j)$ remain and lead to
\begin{equation} 
\label{zeta^k} 
\moy{E^{(k)}} = \frac{Q^2}{2} \Big(
\frac{1}{L^3}\sum_{\substack{\vk\in\KK\\\vk\neq0}} \TF{\phi}(\vk) -
\frac{2\alpha}{\sqrt{\pi}} \Big) = \frac{Q^2}{2}\zeta^{(k)},
\end{equation} 
where the second equality defines $\zeta^{(k)}$.  The average
real-space energy of a single ion of charge~$q_i$ in periodic random
systems is
\begin{equation} 
\label{16} 
\moy{E_i^{(r)}} = \frac{q_i^2}{2}\Big(
\sum_{\substack{\vn\in\ZZ^3\\\vn\neq0}} \psi(\vect{n} L) -
\frac{1}{L^3}\int_{\mathbb{R}^3}\psi(\ver)\,\dd^3\ver \Big)
\end{equation} 
where the first term is the sum of the direct interactions of the ion
with all its periodic images, while the second term corresponds to its
interaction with the uniform background charge density $-q_i/L^3$
provided by the other particles in the system. Since
\begin{equation} \label{int_psi}
\int_{\mathbb{R}^3}\psi(\ver)\,\dd^3\ver
= 4\pi\int_0^\infty r^2\psi(\alpha r)\,\dd r
= \frac{\pi}{\alpha^2},
\end{equation}
we can write the average \emph{total} real-space energy as
\begin{equation} \label{avg_E^r}
\moy{E^{(r)}} = \frac{Q^2}{2}
\Big( \sum_{\substack{\vn\in\ZZ^3\\\vn\neq0}} \psi(\vn L) -
\frac{\pi}{\alpha^2 L^3} \Big) = \frac{Q^2}{2}\zeta^{(r)}, 
\end{equation} 
which defines $\zeta^{(r)}$. The second term in $\zeta^{(r)}$ is, not
surprisingly, identical to the energy~$E^{(n)}$ defined
in~\eqref{E^n}.  Notice that the above result for $\moy{E^{(r)}}$ may
also be obtained by splitting \eqref{U^r} into self ($i=j$) and
interaction terms, and using for the latter $\sum_{j\neq i} q_j =
-q_i$ which follows from the electro-neutrality condition. The
expression of the factor $\zeta = \zeta^{(r)} + \zeta^{(k)}$ is
therefore
\begin{equation} 
\label{zeta}
\zeta = 
\Big( \sum_{\substack{\vn\in\ZZ^3\\\vn\neq0}} 
  \psi(\vn L) - \frac{\pi}{\alpha^2 L^3} 
\Big) +
\Big(
  \frac{1}{L^3}\sum_{\substack{\vk\in\KK\\\vk\neq0}} \TF{\phi}(\vk) - 
  \frac{2\alpha}{\sqrt{\pi}} 
\Big).
\end{equation} 

Eq. \eqref{zeta} can be computed for a number of different box
geometries\cite{Brush}.  For a cubic simulation box of size~$L$, it
yields\cite{Nijboer,Darden}
\begin{equation}  \notag
\zeta\simeq -2.837297479480619610825442578061/L.
\end{equation}

The above calculation shows that, when a charged system is simulated
using PBC, the electrostatic energy~\eqref{Ewald_sum} contains an
additional contribution $Q^2\zeta/2$. The existence of this Madelung
self-energy can be made more apparent in the Ewald formula for~$E$, as
shown in Appendix~\ref{AA}.

\subsection{Madelung cut-off error correction terms}

The Ewald sums~\eqref{U^r} and \eqref{E^k} are necessarily truncated
when evaluated in a simulation. These truncations introduce systematic
cut-off errors in the total energy, because the Madelung self-energies
of the ions are then not fully accounted for. This systematic error is
typically of the same order of magnitude, or even larger, than the
fluctuating error, due to the use of cut-offs, in the Ewald pair
interaction energy~\cite{KolPer,WH}. Note, that no similar systematic
error affects the electrostatic forces, because the Madelung energy
does not depend on the position of the ion. 

Fortunately, it is easy to suppress the systematic bias in the
computed energies. We simply have to add the cut-off correction
\begin{equation} 
\label{U^k_cut} 
E_\mathrm{cut}^{(k)} = \frac{Q^2}{2L^3} \sum_{\substack{\vk\in\KK\\
k>k_\mathrm{cut}}} \TF{\phi}(\vk) 
\end{equation} 
to the computed $k$-space energies, which Kolafa and Perram termed the
\emph{diagonal correction}~\cite{KolPer}. The value of $E_{\rm
  cut}^{(k)}$ does not depend on the configuration and may thus be
computed in advance using a sufficiently large second cut-off $k'_{\rm
  cut}>k>k_\mathrm{cut}$. Using definition \eqref{zeta^k} of~$\zeta^{(k)}$, we can rewrite
\eqref{U^k_cut} as
\begin{equation}
\label{U^k_cut_new} 
  E_\mathrm{cut}^{(k)} = \frac{Q^2}{2} \left(\zeta^{(k)} - \zeta_\mathrm{cut}^{(k)}\right)
\end{equation}
where
\begin{equation}
  \label{zeta^k_cut}
  \zeta_\mathrm{cut}^{(k)} = 
  \frac{1}{L^3} \!\!\!\!\!\!
  \sum_{\substack{\vk\in\KK\\\vk\neq0,\;k<k_\mathrm{cut}}}
  \!\!\!\!\!\!\!\! \TF{\phi}(\vk)
  - \frac{2\alpha}{\sqrt{\pi}}.
\end{equation}

Similarly, if the real-space energies are computed using a cut-off
$r_\mathrm{cut}<L/2$ (minimum image convention), we see from
Eqs.~\eqref{16}, \eqref{int_psi}, and \eqref{avg_E^r}, that the
$r$-space cut-off correction
\begin{equation} 
\label{U^r_cut} 
E_\mathrm{cut}^{(r)} = \frac{Q^2}{2} (\zeta^{(r)} - \zeta_\mathrm{cut}^{(r)})
\end{equation} 
where 
\begin{align} \notag
  \zeta_\mathrm{cut}^{(r)} &=
  -\frac{4\pi}{L^3} \int_{0}^{r_\mathrm{cut}} r^2 \psi(r)\dd r \\ \label{zeta_cut^r}
  &= -\frac{2\pi}{L^3} \left( r_\mathrm{cut}^2 -
    \frac{r_\mathrm{cut}}{\alpha \sqrt{\pi}}
    e^{-\alpha^2 r_\mathrm{cut}^2} 
    - \mathrm{erf}(\alpha r_\mathrm{cut})
    \Big(r_\mathrm{cut}^2-\frac{1}{2\alpha^2}\Big)\right)
\end{align} 
must be applied to the direct space energies.  It is natural that the
correction terms $E^{(k)}_\mathrm{cut}$ and $E^{(r)}_\mathrm{cut}$ are
made up of the exact Madelung energies, minus the average Madelung
energies of the ions as obtained from a calculation with direct and
reciprocal space cut-offs $r_\mathrm{cut}$ and $k_\mathrm{cut}$.

Adding \eqref{U^k_cut_new} to \eqref{U^r_cut} and using~\eqref{zeta}, the two cut-off corrections
can be combined into a single expression
\begin{equation} 
\label{U_cut} 
E_\mathrm{cut} =
E_\mathrm{cut}^{(r)} + E_\mathrm{cut}^{(k)} = 
\frac{Q^2}{2} \Big( \zeta - \zeta_\mathrm{cut}^{(k)} -
\zeta_\mathrm{cut}^{(r)} \Big).
\end{equation} 

All of these terms can easily be precomputed numerically before the
start of a simulation.  

Correcting the systematic cut-off errors in the energies with the term
$E_\mathrm{cut}$ does improve significantly the accuracy of the
results, especially when working with small cut-offs.  In numerical
tests, however, the direct space cut-off correction
$E_\mathrm{cut}^{(r)}$ has been found to be mostly negligible compared
to the reciprocal space correction $E_\mathrm{cut}^{(k)}$ for all
practical purposes.

\section{Mesh-based Ewald sum}
\label{mesh}

The idea of particle-mesh algorithms is to speed up the calculation of
the reciprocal energy~$E^{(ks)}$ with the help of a
Fast-Fourier-Transform (FFT). To use a FFT, the charge density must be
assigned to points on a regular grid. There are several ways of
discretizing the charge density on a grid, and to get the
electrostatic energy from the Fourier transformed grid. We will use
the P3M method of Hockney and Eastwood (but with the standard Ewald
reciprocal interaction~\eqref{def_phi}), because this method surpasses
in efficiency the other variants of mesh based Ewald sums (PME,
SPME)~\cite{DH}.

For simplicity, we assume the number of grid points $M$ to be
identical in all three directions. Let $h=L/M$ be the spacing between
two adjacent grid points. We denote by $\MM$ the set of all grid
points: $\MM=\{\vm h: \vm\in\ZZ^3 \text{\ and\ } 0\leq m_{x,y,z}<
M\}$.

The mesh based calculation of the reciprocal energy is made in the
following steps:

\subsection{Assign charges to grid points}
The charge density $\rho\M(\ver)$ at a grid point~$\ver$ is computed
via the equation
\begin{equation}
  \label{charge_assignment} 
  \rho\M(\ver) = \int
  U(\ver-\ver')\rho(\ver')\dd\ver', \qquad \ver\in\MM, 
\end{equation} 
where $U(\ver)=h^{-3}W(\ver)$ with $W$ the charge assignment function (the factor $h^{-3}$ ensures merely that $\rho\M(\ver)$ has the dimensions of a density). A charge assignement function is classified according to its order~$P$, \ie{} between how many grid points per coordinate direction each charge is distributed. Typically, one chooses a cardinal B-spline for~$W$, which is a piece-wise polynomial function of weight one. The order $P$ gives the number of sections in the function. In P3M, we only need the Fourier transform of the cardinal B-splines, which are
\begin{equation}
\TF{W}^{(P)}(\vk) = h^3 \left( 
\frac{\sin(k_x h/2)}{k_x h/2} \frac{\sin(k_y h/2)}{k_y h/2} \frac{\sin(k_z h/2)}{k_z h/2}
\right)^P.
\end{equation}

Notice that $\rho\M(\ver) = h^{-3}\sum_i q_i\, W(\ver-\ver_i)$, apart at the
boundaries where the periodicity has to
be properly taken into account.

\subsection{Fourier transform the charge grid}
Compute the finite Fourier transform of the mesh-based charge density
(using the FFT algorithm)
\begin{equation} 
\label{FFT_rho} 
\TF{\rho}\M(\vk) =
h^3\sum_{\ver\in\mathbb{M}}\rho\M(\ver)e^{-i\vk\cdot\ver} = {\rm
  FFT}\{\rho\M\} , \qquad \vk\in\TFM.  
\end{equation} 
Here $\vk$ is a wave vector in the reciprocal mesh $\TFM=\{2\pi \vn
/L:\, \vn\in\ZZ^3, |n_{x,y,z}|<M/2\}$.

We stress that $\TF{\rho}\M(\vk)$ differs from $\TF{\rho}(\vk)$ for
$\vk\in\TFM$, because sampling of the charge density on a grid
introduces errors (see Sec. \ref{mesh_errors}).

\subsection{Solve Poisson equation (in Fourier space)}
The mesh-based electrostatic potential $\Phi\M$ is given by the Poisson
equation, which reduces to a simple multiplication in k-space:
\begin{equation} 
\label{no} 
\TF{\Phi}\M(\vk) = \TF{\rho}\M(\vk) \TF{\phi}(\vk),
\qquad \vk\in\TFM, 
\end{equation} 
with $\TF{\phi}(\vk)$ the Fourier transformed
reciprocal interaction~\eqref{TF_phi}.  However, instead of using
$\TF{\phi}(\vk)$ in the above equation, it is better to introduce an
``influence'' function $\TF{G}(\vk)$. We replace therefore
Eq.~\eqref{no} by 
\begin{equation} 
\label{def_G} 
\TF{\Phi}\M(\vk) =
\TF{\rho}\M(\vk) \TF{G}(\vk), \qquad \vk\in\TFM.  
\end{equation} 
where $\TF{G}(\vk)$ is determined by the condition that it leads to
the smallest possible errors in the computed energies (on average for
uncorrelated random charge distributions). $\TF{G}(\vk)$ will be
determined later (see Eq.~\eqref{opt_G}); it can be computed once and
for all at the beginning of a simulation since it depends only on the
mesh size and the charge assignment function. $\TF{G}(\vk)$ plays
basically the same role as the reciprocal interaction
$\TF{\phi}(\vk)$, except that it is tuned to minimize a well defined
error functional in $\TF{\rho}\M(\vk)$. We stress that $\TF{G}(\vk)$
is defined only for $\vk\in\TFM$ (we dropped the subscript $\rm M$ on
the influence function to alleviate the notation). The idea of
optimizing $\TF{G}(\vk)$, which is a key-point of the P3M algorithm,
ensures that the mesh based calculation of the reciprocal energy gives
the best possible results\cite{HE}

\subsection{Get total reciprocal electrostatic energy}
Expression~\eqref{E^k} is approximated on the mesh by
\begin{equation} 
  \label{E^K_NBI} 
  E^{(ks)}_\mathrm{\PPPM} = \frac{1}{2L^3}\sum_{\substack{\vk\in\TFM\\
      \vk\neq0}}|\TF{\rho}\M(\vk)|^2 \TF{G}(\vk).  
\end{equation} 
The total reciprocal energy follows from subtracting the self-energies
from the above quantity: $E^{(k)}_\mathrm{\PPPM} =
E^{(ks)}_\mathrm{\PPPM} - E^{(s)}$.

\subsection{Electrostatic energy of individual charges (optional)}
If the reciprocal energy of each individual particle is needed (and
not only their sum as in step 4), the potential mesh must be
transformed back to real space via an inverse FFT, \ie{}
\begin{equation} 
\label{Phi_M(r)}
\Phi\M(\ver_m) =
\frac{1}{L^3}\sum_{\vk\in\TFM}\TF{\phi}\M(\vk)e^{i\vk\cdot\ver_m} =
\mathrm{FFT}^{-1}\{\TF{\phi}\M\}.  
\end{equation} 
The mesh-based potential is then mapped back to the particle positions
using the same charge assignment function:
\begin{equation} 
\label{Phi(r)} 
\Phi(\ver) :=
\sum_{\ver_m\in\MM_\mathrm{p}} W(\ver-\ver_m) \Phi\M(\ver_m).  
\end{equation} 
In this equation, $\MM_\mathrm{p} = \{\vm
h:\vm\in\ZZ^3\}$ is the mesh extended by periodicity to all space, and
$\Phi\M(\ver)$ is assumed to be periodic (with period $L$). The interpretation of Eq.~\eqref{Phi(r)} is the following: due to the discretization each particle is replaced by several ``sub-particles'' which are located at the surrounding mesh points and carry the fraction $W(\ver-\ver_m)$ of the charge of the original particle. The potential at the position of the original particle is given by the sum of the charge fraction times the potential at each mesh points.
The reciprocal electrostatic energy of the $i^\mathrm{th}$ particle is then
$q_i\Phi(\ver_i)/2$, and the total reciprocal energy (including
self-energies) is the sum
\begin{equation} \label{14}
E^{(ks)}_{\PPPM} = \frac{1}{2}\int_V \rho(\ver)\Phi(\ver)\dd\ver =
\frac{1}{2}\sum_i q_i\, \Phi(\ver_i).  
\end{equation} 
This formula gives the same result for the total energy as
Eq.~\eqref{E^K_NBI}. A mathematical proof of the equivalence is given
in Appendix~\ref{AppB}.

\section{Analysis of discretization errors}
\label{mesh_errors}

If the fast Fourier transform has the benefit of speed, it has the
drawback of introducing errors in the $k$-space spectrum of the charge
density: $\TF{\rho}\M(\vk)$ differs from the true Fourier
transform~\eqref{def_TF_rho}(times a trivial factor $\TF{U}(\vk)$)
because of the discretization on a finite grid.

The difference is two-fold. Firstly, $\TF{\rho}(\vk)$ is defined for
any vector in the full k-space $\mathbb{K}$, whereas
$\TF{\rho}\M(\vk)$ is defined only for $\vk\in\TFM$, \ie{} in the
first Brillouin zone. This is a first natural consequence of
discretization: if the grid spacing is $h$, it necessarily introduces
a cut-off $|k_{x,y,z}|<\pi/h$ in $k$-space. Secondly, the act of
sampling the charge density at grid points, which is mathematically
embodied in Eq.~\eqref{FFT_rho} by the presence of a discrete Fourier
transform instead of a continuous FT, introduces \emph{aliasing}
errors. While a continuous FT would simply transform the convolution
Eq.~\eqref{charge_assignment} into
\begin{equation} 
\mathrm{ FT}\{\rho\M\}(\vk)=\TF{U}(\vk)\TF{\rho}(\vk), \quad \vk\in\KK, 
\end{equation}
the finite Fourier transform results in (see proof in
Appendix~\ref{AppA}) 
\begin{equation} 
\label{SAMPLING_THM} 
\TF{\rho}\M(\vk) = {\rm
  FFT}\{\rho\M\}(\vk) = 
\sum_{\vm\in\ZZ^3}\TF{U}(\vk+\vm
k_g)\TF{\rho}(\vk+\vm k_g), \quad\vk\in\TFM,  
\end{equation} 
where $k_g=2\pi/h$. The sum over $\vm$
shows that spurious contributions from high frequencies of the full
spectrum $\TF{U}(\vk)\TF{\rho}(\vk)$ are introduced into the first
Brillouin zone~$\TFM$. These unwanted copies of the other Brillouin
zones into the first one are known as aliasing errors~\cite{HE}.

To avoid aliasing errors, the spectrum needs to be entirely contained
within the first Brillouin zone. Since $\TF{\rho}(\vk)$ may contain
arbitrary high frequencies, this can only be achieved by choosing
$U(\vk)$ to be a low-pass filter satisfying $\TF{U}(\vk)=0$ for
$\vk\in\KK\setminus\TFM$. But the charge assignment function would
then have a compact support in $k$-space, and hence an infinite
support in $r$-space. This is not acceptable, as it would require the
grid to have an infinite extension. The need to keep the charge assignment function
local in $r$-space means that $\TF{U}(\vk)$ cannot be a perfect low
pass filter. Aliasing errors are therefore unavoidable, and the impact
of these errors must be minimized, by choosing a good compromise for
the charge assignment function and optimizing the influence function. The influence function
can indeed compensate partially for the aliasing errors, because the
spectrum of~$\TF{U}(\vk)$ is known exactly at all frequencies.

The error in reciprocal energy, for a given configuration $\rho(\ver)$
of the charges, is defined by the difference 
\begin{equation} 
\label{def_Delta_E^k}
\Delta E^{(k)} = E^{(k)}_{\PPPM} - E^{(k)} 
\end{equation} 
where $E^{(k)}$ is the exact reciprocal energy (see \eqref{E^k} and
\eqref{E^s}).  The above analysis of discretization errors results in
the explicit formula for this error
\begin{equation} 
\label{error} 
\Delta E^{(k)} =
\frac{1}{2L^3}\sum_{\substack{\vk\in\TFM\\
\vk\neq0}}|\TF{\rho}\M(\vk)|^2 \TF{G}(\vk) - \frac{1}{2L^3}\sum_{\substack{\vk\in\KK\\
\vk\neq0}}|\TF{\rho}(\vk)|^2
\TF{\phi}(\vk), 
\end{equation} 
where $\TF{\rho}\M(\vk)$ is given by~\eqref{SAMPLING_THM}. The error
$\Delta E^{(k)}$ is due to the finite resolution $h$ offered by the
mesh. The finiteness of $h$ introduces the cut-off $\pi/h$ in $k$-space
($\vk\in\TFM$) and causes aliasing errors ($\TF{\rho}\M(\vk) \neq
\TF{\rho}(\vk)\TF{U}(\vk)$) that cannot be entirely eliminated by the
charge assignment function.

\section{Optimization of the P3M algorithm}
\label{p3m_optimization}

We derive in this section the influence function $\TF{G}(\vk)$ that
minimizes the error~\eqref{error} on average for uncorrelated systems,
and give a formula for the associated RMS errors. The average over
random systems is denoted by angular brackets, as in Sec.~\ref{Ewald_correction}.

Notice that the assumption of the absence of correlations is never
satisfied in practice (even for uniform systems because negative
charges tend to cluster around positive charges and vice-versa). The
error estimate proves however to predict quite accurately the error in
real systems with correlations, notably in liquids where the pair
distribution function $g(r)$ decays rapidly to one.

\subsection{Shift in the energies to avoid systematic errors}	\label{S<DeltaU>}

The P3M energies~\eqref{E^K_NBI} contain in general systematic errors,
\ie{} $\moy{\Delta E^{(k)}} \neq 0$, because the Madelung energies
of the ions obtained in the mesh calculation contain cut-off and
aliasing errors. The average error 
\begin{equation} 
\label{def_K} 
K = \moy{\Delta
  E^{(k)}} = \moy{E_{\PPPM}^{(k)}} - \moy{E^{(k)}} 
\end{equation} 
is a constant that must be \emph{subtracted} from the P3M energies, to
ensure that the energies are right on average. The corrected P3M
energies are thus obtained by applying a constant shift to the
original P3M energies:
\begin{equation}
\label{P3M_U^k} E_{\PPPM, \text{corr}}^{(k)} = E^{(k)}_\mathrm{\PPPM} - K, 
\end{equation} 
where the constant $K$ depends on the various P3M parameters like mesh
size, charge assignment order (CAO) and Ewald splitting parameter.

Let us determine analytically the constant~\eqref{def_K}. Writing it
as $K = \moy{E_{\PPPM}^{(k)}} - \moy{E^{(k)}}$, we can use the result
\eqref{zeta^k} for $\moy{E^{(k)}}$: it is nothing but
$Q^2\zeta^{(k)}/2$, \ie{} the $k$-space Madelung energies of the
ions. The other term $\moy{E_{\PPPM}^{(k)}}$ can be calculated in the
same way as~\eqref{Calc_zeta^k}. Using \eqref{E^K_NBI},
\eqref{SAMPLING_THM} and \eqref{E^s}, we find
\begin{equation} 
\label{<U^k_P3M>}
\moy{E^{(k)}_{\PPPM}} = \frac{Q^2}{2}
\Big( \frac{1}{L^3}
\sum_{\substack{\vk\in\TFM\\\vk\neq0}}
\TF{G}(\vk)\sum_{\vm\in\ZZ^3}\TF{U}^2(\vk+k_g\vm) - \frac{2\alpha}{\sqrt{\pi}}\Big) = \frac{Q^2}{2}
\zeta^{(k)}_{\PPPM},
\end{equation} 
which defines $\zeta^{(k)}_{\PPPM}$.
The result~\eqref{<U^k_P3M>} can be interpreted as the average
$k$-space Madelung energies of the ions \emph{as obtained from the mesh
  calculation}, \ie{} including cut-off and aliasing errors. The
explicit expression of the correction constant~\eqref{def_K} is thus
\begin{equation}
\label{K_explicit} 
{ K = \frac{Q^2}{2} \left( \zeta^{(k)}_{\PPPM} - \zeta^{(k)} \right) =
  \frac{Q^2}{2L^3} \Big( \sum_{\substack{\vk\in\TFM\\
\vk\neq0}} \TF{G}(\vk)\sum_{\vm\in\ZZ^3}\TF{U}^2(\vk+k_g\vm) -
  \sum_{\substack{\vk\in\KK\\
\vk\neq0}} \TF{\phi}(\vk) \Big) } 
\end{equation} 
In the last sum in~\eqref{K_explicit}, the terms with $|k_{x,y,z}|>\pi/h$ are
equivalent to the $k$-space cut-off correction defined
in~\eqref{U^k_cut}. These terms compensate for the fact that the
Madelung energies of the ions are underestimated in the mesh calculation
because of the cut-off $\pi/h$ introduced by the finite size of the
mesh. The remaining terms in~\eqref{K_explicit} compensate, on
average, the aliasing errors that affect the Madelung energies of the
ions obtained from the mesh calculation.

Notice that the two correction terms $E^{(r)}_\mathrm{cut}$ and $-K$ can
be combined together in the simple expression 
\begin{equation} 
\label{39}
E^\mathrm{cut}_\mathrm{P3M} = E^{(r)}_\mathrm{cut} - K = \frac{Q^2}{2} \Big( \zeta -
\zeta^{(k)}_{\PPPM} - \zeta^{(r)}_{\rm cut} \Big)
\end{equation} 
where $\zeta$ is defined by~\eqref{zeta} and $\zeta^{(r)}_{\rm cut}$ is given in~\eqref{zeta_cut^r}. We stress that $E^\mathrm{cut}_\mathrm{P3M}$
has the same structure as the correction term~\eqref{U_cut} for
truncated Ewald sums. The difference lies in the replacement of $\zeta_{\rm cut}^{(k)}$ by the quantity $\zeta_{\PPPM}^{(k)}$ defined in~\eqref{<U^k_P3M>}, which accounts for both the cut-off and aliasing errors that affect the reciprocal energies computed on the mesh.

In summary, the final formula for computing the total electrostatic energy with the P3M algorithm is
\begin{equation}	\label{TOTAL_ENERGY_P3M}
  \begin{split}
    E & \approx E^{(r)} & \mathrm{[eq. \eqref{U^r}]}\\
    & + E^{(ks)}_{\PPPM} & \mathrm{[eq. \eqref{E^K_NBI}]}\\
    & - E^{(s)} & \mathrm{[eq. \eqref{E^s}]}\\
    & + E^{(d)} & \mathrm{[eq. \eqref{E^d}]}\\
    & + E^{(n)} & \mathrm{[eq. \eqref{E^n}]}\\
    & + E^\mathrm{cut}_\mathrm{P3M} & \mathrm{[eq. \eqref{39}]}\\
  \end{split}
\end{equation}
The correction term~$E^\mathrm{cut}_\mathrm{P3M}$ is necessary to compensate on average for systematic errors in the mesh calculation. It can be computed once for all before the start of a simulation, since it depends only on the size of the simulation box, the size of a mesh cell, the charge assignment function and the influence function.

\subsection{RMS error estimate for energy}

The result~\eqref{error} is an exact measure of the error in the P3M
energies for a given configuration $\rho(\ver)$ of the particles. Let
us average this expression over all possible positions of the
particles to get a useful overall measure of the accuracy of the
algorithm. The RMS error of the corrected P3M energies is, by definition,
\begin{equation} 
\label{25} 
(\Delta E^{(k)}_{\rm
  RMS})^2 := \moy{(E_{\PPPM, \text{corr}}^{(k)}-E^{(k)})^2} =
\moy{(\Delta E^{(k)} -K)^2}
\end{equation}
where we used \eqref{def_Delta_E^k} and \eqref{P3M_U^k}. We can
isolate in $\Delta E^{(k)}$ ``interaction'' terms ($i\neq j$) from
self terms ($i=j$):
\begin{equation} 
\label{27} 
(\Delta E^{(k)}_\mathrm{RMS})^2 = 
\moy{\Big(\Delta E_\mathrm{int}^{(k)}+\Delta E_\mathrm{self}^{(k)} - K\Big)^2}.  
\end{equation} 
We recall from Sec.\ref{Ewald_correction} that the interaction terms vanish on
average for random systems: $\moy{\Delta E_\mathrm{int}^{(k)}}=0$. The
correlation
\begin{equation} 
\label{approx1} 
\moy{\Delta E_\mathrm{self}^{(k)}\cdot\Delta E_\mathrm{int}^{(k)}} = 0 
\end{equation} 
vanishes as well for the same reason (this is due to the fact that the
average Ewald interaction energy between a fixed particle~$i$ and a
particle $j\neq i$ is zero, see Appendix~\ref{AA}).  Eq.~\eqref{27}
reduces therefore to
\begin{equation} 
\label{30} 
(\Delta E^{(k)}_\mathrm{RMS})^2 = \moy{(\Delta E_\mathrm{int}^{(k)})^2}
+ \moy{(\Delta E_\mathrm{self}^{(k)}-K)^2},  
\end{equation} 
where the first term accounts for fluctuating errors in the
interactions energies, and the second term accounts for fluctuating
errors in the corrected Madelung self-energies of the ions. Since the
latter term may be written as
\begin{equation}
 \moy{(\Delta E_\mathrm{self}^{(k)}-K)^2} = \moy{(\Delta E_\mathrm{self}^{(k)})^2} - K^2,
\end{equation}
we remark that the shift $-K$ derived in the previous section, in
addition to removing the systematic bias in the $k$-space energies,
also reduces the fluctuating errors of the $k$-space self-energies by
an amount $-K^2$.

In the substraction $\Delta E_\mathrm{self}^{(k)}-K$, it can be seen,
from \eqref{error} in which only $i=j$ terms are kept and
\eqref{K_explicit}, that all terms containing $\TF{\phi}(\vk)$ cancel
out, so we have
\begin{multline}
\moy{(\Delta E_\mathrm{self}^{(k)}-K)^2} = 
\left\langle\Big( \frac{1}{2L^3} \sum_{\substack{\vk\in\TFM\\ \vk\neq 0}} \TF{G}(\vk) \sum_i q_i^2
 \Big[\sum_{\vm_1}\sum_{\vm_2}\TF{U}(\vk_{\vm_1})\TF{U}(\vk_{\vm_2}) e^{i k_g (\vm_1-\vm_2)\cdot\ver_i} \right.\\\left.
 - \sum_{\vm} \TF{U}^2(\vk_{\vm})\Big] \Big)^2 \right\rangle
\end{multline}
where we used the symmetry $\TF{U}(-\vk)=\TF{U}(\vk)$ and introduced
the shorthand notation $\vk_{\vm}=\vk+k_g\vm$.  When the square is
expanded, the summation over particles $\sum_i$ becomes a double
summation $\sum_{i,i'}$. All terms with $i'\neq i$ vanish, because
$\moy{\exp(i k_g (\vm_1-\vm_2)\cdot\ver_i)}=\delta_{\vm_1,\vm_2}$,
leaving identical sums over $\vm$ which cancel each other. The
remaining terms $i'=i$ evaluate to
\begin{multline}
\moy{(\Delta E_\mathrm{self}^{(k)}-K)^2} = 
\frac{1}{4L^6} (\sum_i q_i^4) \sum_{\substack{\vk\in\TFM\\ \vk\neq 0}} \sum_{\substack{\vk'\in\TFM\\ \vk'\neq 0}} \TF{G}(\vk) \TF{G}(\vk') \times \\ \times
\Big\{ 
 \sum_{\vm_1}\sum_{\vm_2}\sum_{\vm_3} \TF{U}(\vk_{\vm_1})\TF{U}(\vk_{\vm_2})\TF{U}(\vk'_{\vm_3})\TF{U}(\vk'_{\vm_1-\vm_2+\vm_3}) 
 - \sum_{\vm} \TF{U}^2(\vk_{\vm}) \cdot \sum_{\vm'} \TF{U}^2(\vk'_{\vm'}) \Big\} \\
=  	\label{def_H^2_self}
\frac{1}{4L^3} (\sum_i q_i^4) H^2_\mathrm{self}
\end{multline}
with
\begin{equation}	\label{H^2_self}
 H^2_\mathrm{self} = \frac{1}{L^3}\sum_{\substack{\vk\in\TFM\\ 
  \vk\neq 0}} \sum_{\substack{\vk'\in\TFM\\ 
  \vk'\neq 0}} \TF{G}(\vk) \TF{G}(\vk')
\Big\{ 
 \sum_{\vm_1}\sum_{\vm_2\neq \vm_1}\sum_{\vm_3} \TF{U}(\vk_{\vm_1})\TF{U}(\vk_{\vm_2})\TF{U}(\vk'_{\vm_3})\TF{U}(\vk'_{\vm_1-\vm_2+\vm_3}) \Big\}.
\end{equation}
The fluctuating errors of the Madelung self-energies scale therefore
like $\sum_i q_i^4$ with the valencies of the ions. The prefactor is
somewhat complicated since it involves a double summation over
wave vectors and a triple summation over alias indices
$\vm_1,\vm_2,\vm_3$, but $H^2_\mathrm{self}$ can be evaluated
reasonably fast. The numerical calculation of $H^2_\mathrm{self}$ can be accelerated by taking profit of the symmetries (the sum over $\vk$ can be restricted to only half an octant of the reciprocal mesh), and by skipping inner loops in the triple summation over alias indices if the product of the charge fractions is almost zero.

We calculate now the fluctuations of the errors in the interaction
energies, \ie{} the first term of Eq.~\eqref{30}. That term
reads, using \eqref{27}, \eqref{error}, \eqref{SAMPLING_THM} and
\eqref{def_TF_rho} and keeping only interaction terms:
\begin{multline}	\label{31}
   \moy{(\Delta E_\mathrm{int}^{(k)})^2}
   = \left\langle \Big(
   \frac{1}{2L^3}\sum_{\substack{\vk\in\TFM\\ \vk\neq 0}} \sum_{\substack{i,j\\i\neq j}} q_i q_j 
   \sum_{\vm\in\ZZ^3} e^{i \vk_{\vm}\cdot\ver_i}
   \times \right. \\ \left.
   \times
   \Big[\TF{G}(\vk) \sum_{\vm'\in\ZZ^3} e^{-i \vk_{\vm'}\cdot\ver_j}  \TF{U}(\vk_{\vm}) \TF{U}(\vk_{\vm'})
   - e^{-i \vk_{\vm}\cdot\ver_j}\TF{\phi}(\vk_{\vm}) \Big]
   \Big)^2 \right\rangle.
\end{multline}
The calculation of this average is
straightforward, though somewhat tedious. We find that it reduces to 
\begin{equation} 
\label{32} 
\moy{(\Delta E_\mathrm{int}^{(k)})^2} \simeq
\frac{Q^4}{4L^3} H^2_\mathrm{int}, 
\end{equation} 
where 
\begin{equation} 
\label{def_Q_HE} H^2_\mathrm{int} =
\frac{2}{L^3}\sum_{\substack{\vk\in\TFM\\ 
\vk\neq 0}} \left[\TF{G}^2(\vk)\Big(\sum_{\vm}\TF{U}^2(\vk_{\vm})\Big)^2 -
2 \TF{G}(\vk) \sum_{\vm}\TF{U}^2(\vk_{\vm}) \TF{\phi}(\vk_{\vm})
  + \sum_{\vm}\TF{\phi}^2(\vk_{\vm}) \right].  
\end{equation}
The factor 2 in $H^2_\mathrm{int}$ originates from the fact that each
pair of particles appears twice in the sum over $i$ and $j\neq i$
in~\eqref{31}. Expression~\eqref{def_Q_HE} is the analog for the
energy of the parameter $Q$ introduced by Hockney and Eastwood to
measure the accuracy of the P3M forces~\cite{HE}. Notice that
\eqref{def_Q_HE} is given in real space by
\begin{equation} 
H^2_\mathrm{int} =
\frac{2}{V_\mathrm{cell}}\int_{V_\mathrm{cell}}\dd\ver_1 \int_{L^3}\dd\ver
\, [\phi_{\PPPM}(\ver;\ver_1)-\phi(\ver-\ver_1)]^2 
\end{equation} 
where $\phi_\PPPM(\ver;\ver_1)$ is the reciprocal potential at $\ver$
created by a unit charge located at $\ver_1$, as obtained from the P3M
algorithm. (This potential is given in Fourier space by combining
\eqref{TF_Phi} with \eqref{SAMPLING_THM} in which we set
$\rho(\ver)=\delta(\ver-\ver_1)$.) $H^2_\mathrm{int}$ is hence twice the squared
deviation between the potential $\phi_{\PPPM}$ obtained from the mesh
calculation and the exact reciprocal potential~$\phi$, summed over all
relative positions~$\ver$ within the simulation box, and averaged over
all possible positions of charge~$\ver_1$ in a mesh cell ($V_{\rm
  cell}=h^3$).

Inserting the above results in \eqref{30}, our final expression for
the RMS error of the (corrected) P3M energies is
\begin{equation} 
\label{RMS_error}
\Delta E^{(k)}_\mathrm{RMS} = \frac{\sqrt{Q^4 H^2_\mathrm{int} + (\sum_i q_i^4) H^2_\mathrm{self} }}{2L^{3/2}}
\end{equation}
where $H^2_\mathrm{int}$ and $H^2_\mathrm{self}$ are defined in Eqs.
\eqref{def_Q_HE} and \eqref{H^2_self}. This error depends on the
influence function $\TF{G}(\vk)$. The optimal influence function (the
one that minimizes the error) will be determined in the next section.
The above error estimate, together with the optimal influence
function~\eqref{opt_G} and the constant shift~\eqref{K_explicit} which
must be applied to the P3M energies, constitute the main results of
this paper.

The RMS error~\eqref{RMS_error} displays two different scalings with
the valencies of the ions: $(\sum_i q_i^2)^2$ for errors coming from
pair interactions (such a scaling also governs errors in P3M
forces~\cite{DH2}) and $\sum_i q_i^4$ for errors in Madelung
self-energies. Because of these different scalings, the errors from
pair interactions are expected to dominate in systems with many
charged particles ($Q^4 \gg \sum_i q_i^4)$. Notice that $H^2_{\rm
  self}$ is, roughly speaking, proportional to $(\sum_{\vk}
\TF{G}(\vk))^2$, while $H^2_\mathrm{self}$ scales like $\sum_{\vk}
\TF{G}^2(\vk)$. The errors in the Madelung self-energies increase
therefore more rapidly than the errors in the pair interaction
energies when the Ewald splitting parameter~$\alpha$ (and hence
$\TF{G}(\vk)$) is increased, or when the size of the mesh is
increased. The importance of the two source of errors (fluctuations in
pair interaction energies versus fluctuations in Madelung
self-energies) will be compared in Sec.~\ref{p3m_check} for a test
system with $Q^2=100$.

\subsection{Optimal influence function}	
\label{optimal_influence_function}

We can now determine the optimal influence function $\TF{G}(\vk)$, by
imposing the condition that it minimizes the RMS
error~\eqref{RMS_error}. Since the errors coming from pair P3M
interactions are expected to dominate the self-interaction errors
(except in systems with few particles), we optimize the influence
function only with respect to the pair interactions. Setting
\begin{equation} 
\frac{\delta H^2_\mathrm{int}}{\delta \TF{G}(\vk)} = 0, 
\end{equation}
gives immediately 
\begin{equation} 
\label{opt_G} 
{\TF{G}(\vk) = \frac{\ds\sum_{\vm\in\ZZ^3}\TF{U}^2(\vk_{\vm})\TF{\phi}(\vk_{\vm})}{\ds\Big(\sum_{\vm\in\ZZ^3}\TF{U}^2(\vk_{\vm})\Big)^2}}
\end{equation}
where we recall that the Fourier-transformed reciprocal interaction
$\TF{\phi}(\vk)$ is given by~\eqref{TF_phi}.  An optimization of the
influence function with respect to the \emph{full} RMS error could be
performed, but would require solving a linear system of $M^3$
equations to compute $\TF{G}(\vk)$. The numerical results shown in
Sec.~\ref{p3m_check} will confirm that such a full optimization is not
necessary in typical systems.

Since $\TF{\phi}(\vk)$ decays
exponentially fast, the optimal influence function is given in good
approximation by 
\begin{equation} 
\label{opt_G_approx} 
{\TF{G}(\vk) \simeq \TF{\phi}(\vk) \,
  \frac{\TF{U}^2(\vk)}{\left(\sum_{\vm\in\ZZ^3}\TF{U}^2(\vk_{\vm})\right)^2}}.
\end{equation} 
$\TF{G}(\vk)$ differs thus from $\TF{\phi}(\vk)$ by a factor which
is always less than one. This damping of the interaction compensates
as well as possible for the aliasing errors introduced by the use of a
fast Fourier transform. If $\TF{U}(\vk)$ were a perfect low-pass
filter ($\TF{U}(\vk_{\vect{m}})=0$ if $\vect{m}\neq 0$), no aliasing
error would occur and the influence function would reduce to
$\TF{G}(\vk)=\TF{\phi}(\vk)/\TF{U}^2(\vk)$. This is indeed the result
expected from~\eqref{error} when aliasing errors are absent. The true
optimal influence function~\eqref{opt_G} differs from this simple
expression by contributions from the high-frequency spectrum of $\TF{U}(\vk)$ and reciprocal interaction~\eqref{TF_phi}.

Hockney and Eastwood obtained the following optimal influence function
by minimizing the errors in the forces instead of the energy~\cite{HE}:
\begin{equation} 
\label{opt_G_forces} 
\TF{G}^{(\rm forces)}(\vk) =
\frac{\sum_{\vm}(\vk\cdot\vk_{\vm})\,\TF{U}^2(\vk_{\vm})\TF{\phi}(\vk_{\vm})}{\vk^2
  \left(\sum_{\vm}\TF{U}^2(\vk_{\vm})\right)^2} 
\end{equation} 
Obviously, this function is also given in very good approximation
by~\eqref{opt_G_approx}. This explains why influence functions
\eqref{opt_G}, \eqref{opt_G_approx} and \eqref{opt_G_forces} all give
very similar results when computing energies and forces.

Inserting \eqref{opt_G} into \eqref{def_Q_HE}, we find that the minimal
value of~$H^2_\mathrm{int}$ is 
\begin{equation} 
\label{Q^min_HE} 
\left. H^2_\mathrm{int}\right|_{\rm
    min} = \frac{1}{L^3}\sum_{\substack{\vk\in\TFM\\ 
\vk\neq 0}} \left[\sum_{\vm\in\ZZ^3}\TF{\phi}^2(\vk_{\vm})
    - \left( \frac{\sum_{\vm}\TF{U}^2(\vk_{\vm})\TF{\phi}(\vk_{\vm})}
    {\sum_{\vm}\TF{U}^2(\vk_{\vm})} \right)^2 \right].  
\end{equation} 
This is the expression of $H^2_\mathrm{int}$ to be used in the RMS error
estimate~\eqref{RMS_error} when the P3M algorithm is optimized to
yield the smallest possible errors in the pair interaction energies.

We recall that the errors in the P3M energies originate from aliasing
effects (due to the sampling on a grid) and truncation errors (due to
the fact that the reciprocal mesh contains only a finite number of
wave vectors). The truncation error can only be reduced by choosing a
larger mesh or by using a reciprocal interaction with a faster decay
in $k$-space, whereas the aliasing errors may be reduced by increasing
the order of the charge assignment function (up to the maximum order
allowed by the size of the mesh). The intrinsic truncation error of a
given mesh and reciprocal interaction can be obtained by assuming
$\TF{U}(\vk)$ in~\eqref{Q^min_HE} to be a perfect low-pass filter:
\begin{equation} 
\label{Q^min_cut-off} 
{ \left. H^2_\mathrm{int, cut-off}\right|_\mathrm{min} =
  \frac{1}{L^3}\sum_{\substack{\vk\in\TFM\\ 
\vk\neq 0}} \left[\sum_{\vm\in\ZZ^3}\TF{\phi}^2(\vk_{\vm}) - 
\TF{\phi}^2(\vk) \right]}.  
\end{equation} 
By inserting this formula in~\eqref{RMS_error}, we
get an estimate of the intrinsic RMS cut-off error in $k$-space, caused
by the finite number of wave vectors in the reciprocal mesh. The RMS
error associated with~\eqref{Q^min_cut-off} depends only on the size of
the mesh and on the choice of the reciprocal interaction, \ie{}
Ewald parameter $\alpha$ if the standard form~\eqref{def_phi} is used.

\section{Numerical check of accuracy}
\label{p3m_check}

In this section, we test the analytical results (optimal influence
function, energy shift~$E_{\PPPM}^{\rm cut}$, RMS error estimate) derived in the
previous section.  We do this by comparing the P3M energies with the
exact energies calculated in a specific random system.  In the
following, all dimensions are given in terms of the arbitrary length
unit $\cal L$ and charge unit $\cal C$. In particular, energies and
energy errors are given in units of ${\cal L}^2/{\cal C}$.  We choose
the same test system as the one defined in Appendix~D of Deserno and
Holm\cite{DH}: 100 particles randomly distributed within a cubic box
of length $L=10 {\cal L}$, half of them carry a positive, the other
half a negative unit charge.  The statistical average $\moy{\cdots}$
is calculated by averaging over at least 100 different configurations
of this test system (these configurations are determined by using the
same random number generator as in Deserno and Holm\cite{DH}). Well
converged Ewald sums (in metallic boundary conditions) were used to compute the exact energies of the
test systems. The first three systems have energies $-15.43059$,
$-15.26641$ and $-15.59147$ respectively, values that are all quite
close to the Madelung energies of the ions $Q^2 \zeta / 2 \simeq
-14.187$.

The P3M energies of the test systems were computed with various mesh
sizes ($M=4,8,16,32$), a real-space cut-off $r_\mathrm{cut}=4.95$, and
different orders of the charge assignment function (from 1 to 7). Our
calculations show that using the energy-optimized influence
function~\eqref{opt_G}, instead of the force-optimized influence
function~\eqref{opt_G_forces}, leaves the energies almost unchanged.
(A slight improvement in accuracy appears only when the aliasing error
are at their maximum, namely for a charge assignment order of~1 and
large values of~$\alpha$.) This behavior could have been expected,
since both influence functions are almost equivalent to the simple
formula~\eqref{opt_G_approx}.

\begin{figure}[ht]
\begin{center}
\includegraphics[scale=0.5]{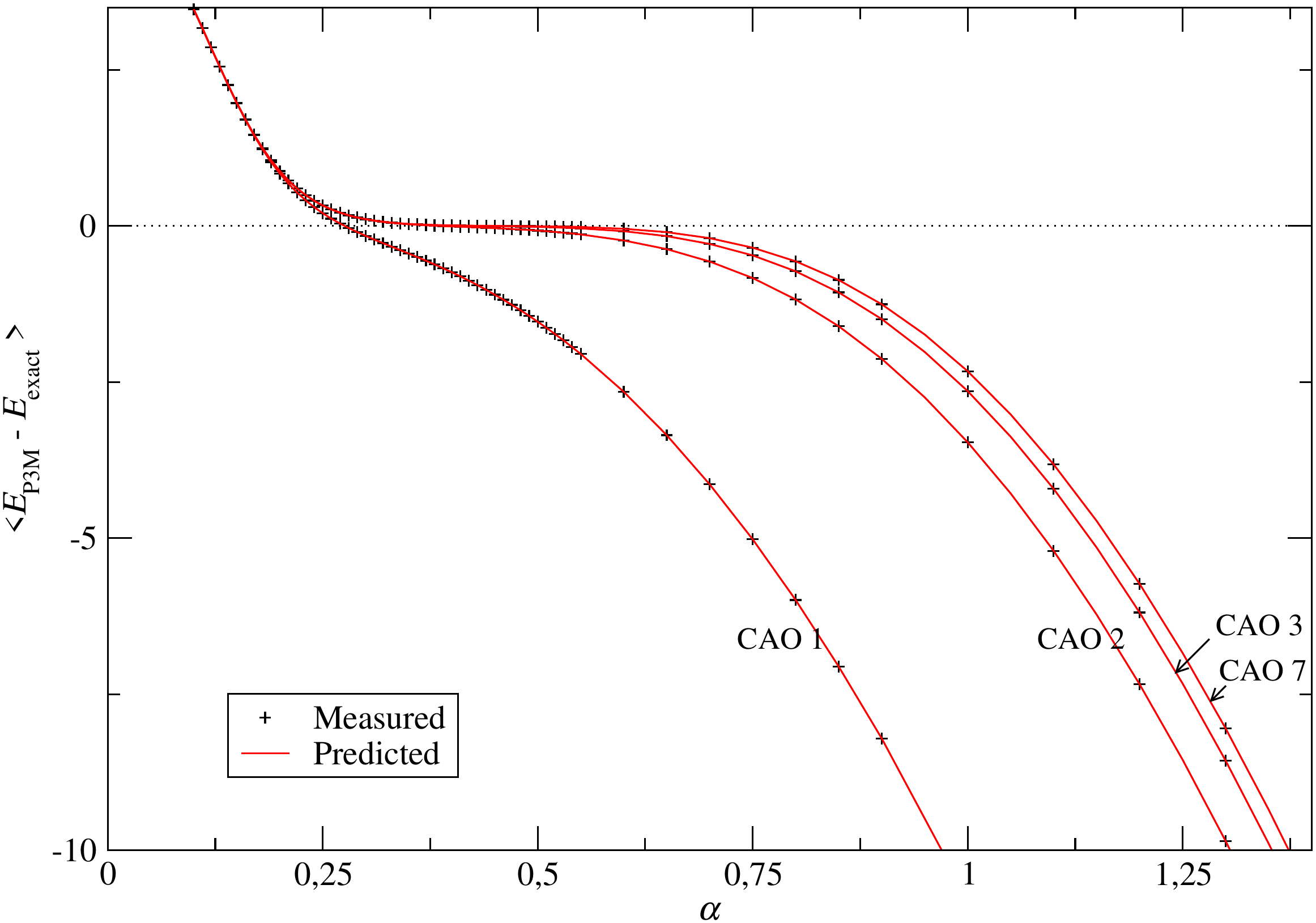}
\caption{\label{Fig1} Comparison between the measured systematic error
  of the uncorrected P3M energies (crosses) and the theoretical
  prediction~$-E_\mathrm{P3M}^\mathrm{cut}$ (solid lines) as a function of Ewald
  parameter~$\alpha$. The average is performed over 1000 test systems
  consisting in 100 charges located at random in a box of size
  $L=10$. The mesh has size~$M=8$ and the real-space cut-off is 
  $r_\mathrm{cut}=4.95$.}
\end{center}
\end{figure}

We compare in Fig.~\ref{Fig1} the measured systematic error
$\moy{E_{\PPPM}-E_{\rm exact}}$ of the uncorrected P3M
energies [Eq.~\eqref{TOTAL_ENERGY_P3M} without term $E_\mathrm{P3M}^\mathrm{cut}$], for CAO's
ranging from 1 to 7, to the expected bias $-E_\mathrm{P3M}^\mathrm{cut}$. The agreement is perfect for all CAO's and for all values of Ewald's splitting parameter~$\alpha$. The energy shift $E_\mathrm{P3M}^\mathrm{cut}$ in Eq.~\eqref{TOTAL_ENERGY_P3M} removes therefore entirely the systematic error, as it should.

Fig.~1 illustrates that the systematic errors in the uncorrected
energies, which are due to cut-off and aliasing errors in the Madelung
self-energies of the ions, have two different contributions of
opposite sign. At small values of $\alpha$, the $r$-space cut-off error
dominates and leads to an overestimation of the energy because the
negative interaction energy of an ion with the neutralizing background
charge provided by the other particles is not fully taken into
account. The cut-off correction \eqref{U^r_cut} derived in
Sec.~\ref{Ewald_correction} does compensate very well for this effect.
At large values of $\alpha$, $k$-space cut-off and aliasing errors
dominate, and lead to an underestimation of the Madelung self-energies
(expression~\eqref{K_explicit} is indeed always negative).
  
Since the systematic error $\langle E^{(k)}_{\PPPM}-E^{(k)}_{\rm
  exact} \rangle$ in the reciprocal energies arise solely from self
terms (the Ewald interaction between a pair of particles is zero on
average), this error can alternatively, and more efficiently, be
measured by computing the P3M energy of a system made up of a single
ion in the box, averaging that energy over different positions of the
particle relative to the mesh. To restore electro-neutrality, the
interaction energy~\eqref{E^n} with the (implicit) neutralizing
background must of course be taken into account before comparing the
result with the exact Madelung self-energy $q^2\zeta/2$ of the ion.
This method allows one to measure very rapidly the reciprocal
contribution to the average error in the Madelung self-energies of the
ions.  We stress that the numerical results shown in Fig.~1 can easily
be transposed to any cubic system with an arbitrary number of ions
since the energy shift $E_\mathrm{P3M}^\mathrm{cut}$ scales merely as $Q^2/L$.

%

\begin{figure}[h]
\begin{center}
\includegraphics[scale=0.5]{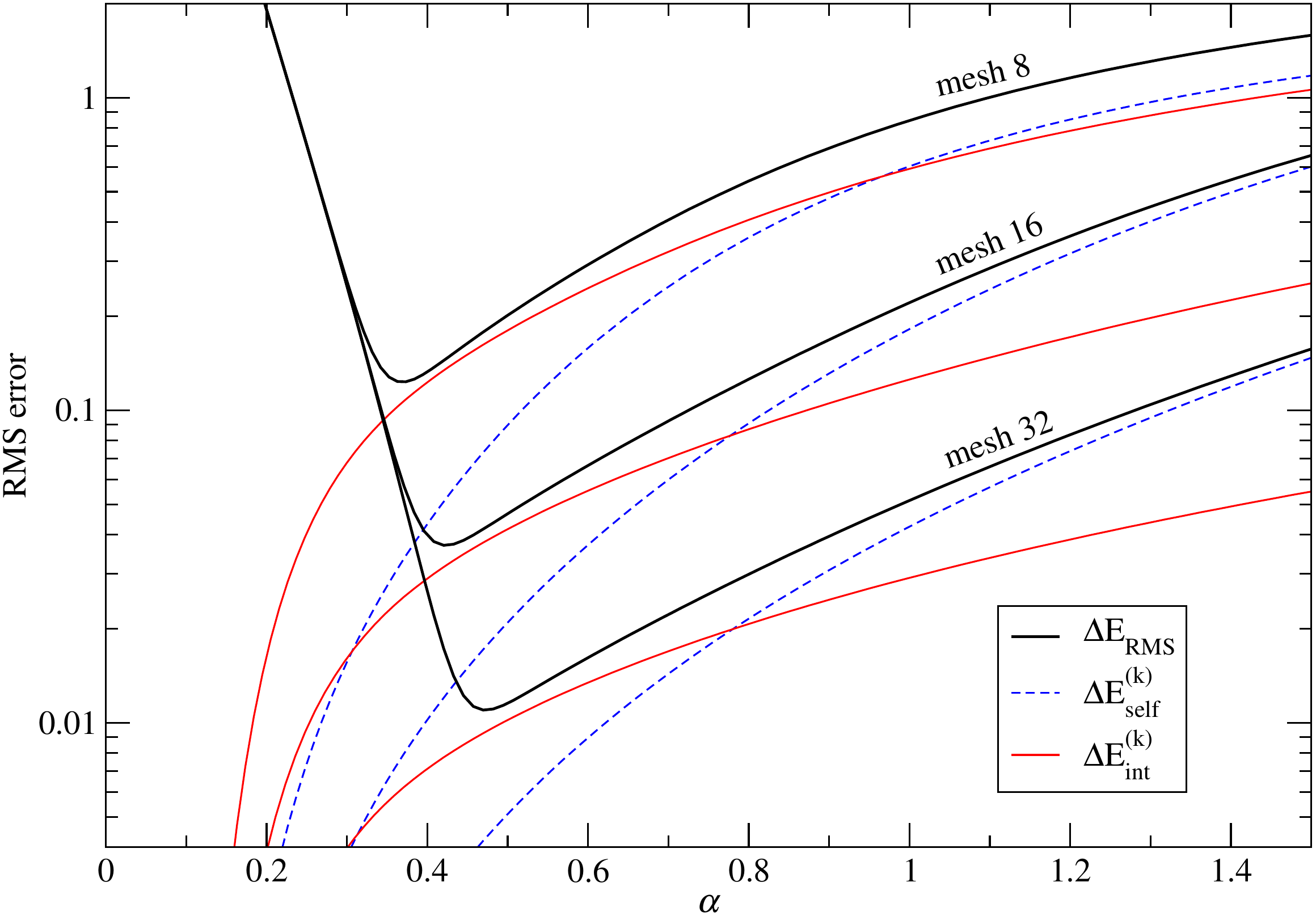}
\caption{\label{Fig2} Theoretical predictions for the RMS error of the
  corrected P3M energies for CAO 2 and three different mesh sizes
  [thick solid lines], for the same system and real-space cut-off as in
  Fig.~\ref{Fig1}. The two contributions which make up the total
  $k$-space error are also shown independently: RMS error in the pair
  P3M interaction energies (Eq.~\eqref{32}, thin solid line) and RMS
  error in the Madelung self-energies (Eq.~\eqref{def_H^2_self},
  dashed line).}
\end{center}
\end{figure}

Having validated the energy shift~\eqref{39}, we test now the accuracy
of the RMS error estimate~\eqref{RMS_error}. We show in
Fig.~\ref{Fig2} the theoretical predictions for the RMS error of the
corrected P3M energies for different mesh sizes (thick solid lines),
at fixed CAO~2. The dominant error at small values of~$\alpha$ comes
from the truncation in the real-space calculation, while $k$-space
cut-off and aliasing errors dominate at large values of $\alpha$. The
plot shows also separately the contribution $\Delta E^{(k)}_{\rm
  self}$, which accounts for fluctuating errors in the $k$-space
Madelung self-energies, and the contribution $\Delta E^{(k)}_{\rm
  int}$ which accounts for fluctuating errors in the P3M pair
interaction energies. Near the optimal value of $\alpha$, the error
$\Delta E^{(k)}_\mathrm{int}$ dominates slightly $\Delta E^{(k)}_{\rm
  self}$ by half an order of magnitude. This validates the use of the
optimal influence function~\eqref{opt_G}, which was designed to
minimize errors in the pair P3M interaction energies only. Notice that
$\Delta E^{(k)}_\mathrm{self}$ overcomes $\Delta E^{(k)}_\mathrm{int}$ at
large values of $\alpha$, in agreement with the scaling with~$\alpha$
discussed in Sec.~\ref{p3m_optimization}. The errors in Madelung
self-energies must therefore by included to predict correctly the full
RMS error curve in our test system with 100 charged particles, but
they are expected to become negligible when the number of ions is
increased above a few hundred.

The predicted RMS errors agree very well with the measured RMS errors,
as shown in Fig.~3. The small deviations at low values of~$\alpha$ are
due to a loss of accuracy of Kolafa and Perram's $r$-space error
estimate~\eqref{RMS_U^r}, and to the fact that this error estimate does not take into
account the improvement in accuracy brought by the new cut-off
correction term~\eqref{U^r_cut}.
In the regime where the dominant
error comes from the $k$-space calculation, the agreement with our RMS
error estimate is excellent, especially at high values of the charge
assignment order. The errors in the $k$-space
calculation are caused by truncation and aliasing effects. The
aliasing errors can be reduced by increasing the charge assignment
order, but the accuracy cannot go below the minimum $k$-space cut-off
error~\eqref{Q^min_cut-off} (dashed curve in Fig.~3), which is
intrinsic to the mesh size and choice of reciprocal interaction.

\begin{figure}[h]
\begin{center}
\includegraphics[scale=0.5]{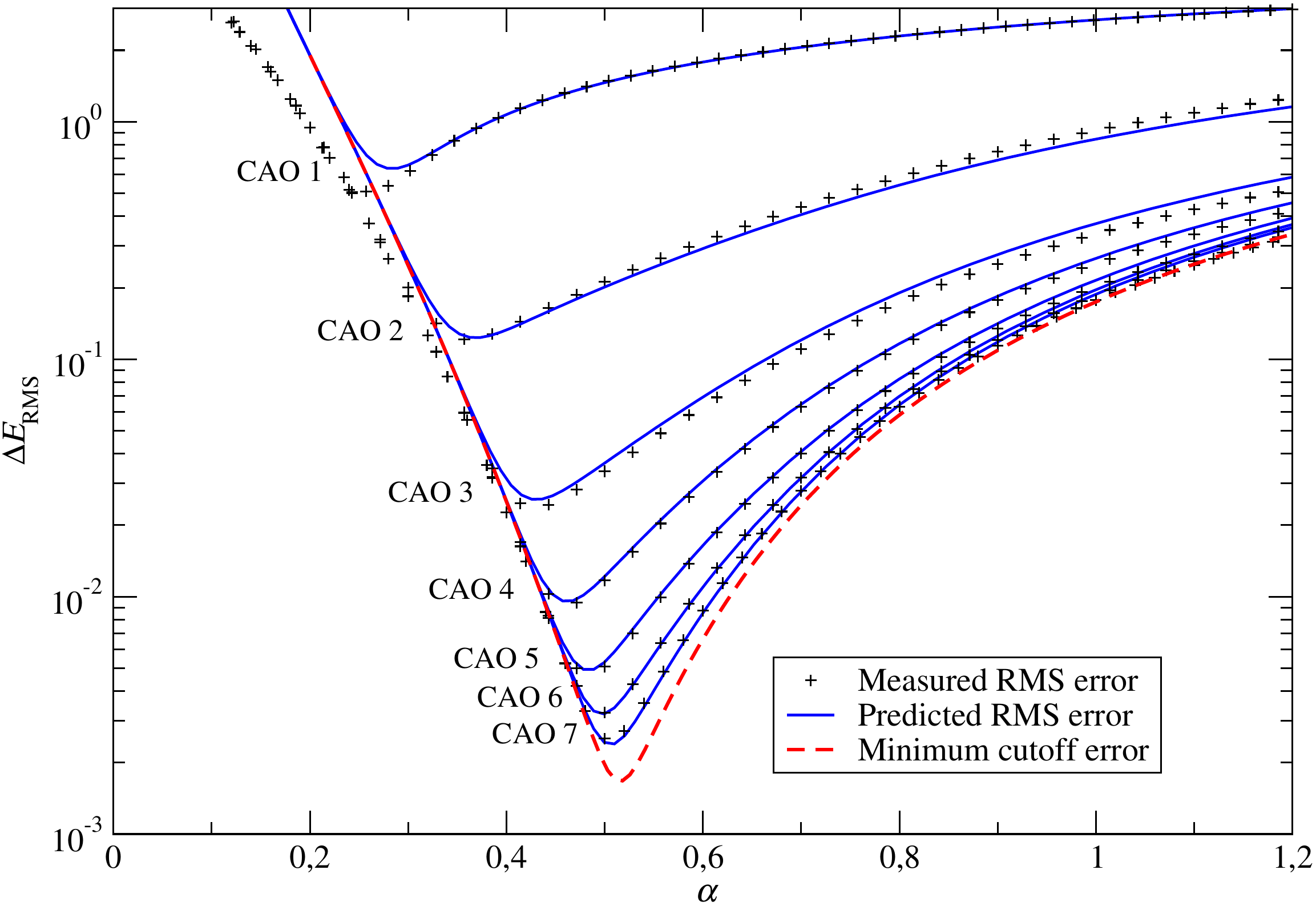}
\caption{\label{Fig3} Comparison between the measured (crosses) and
  predicted (solid lines) RMS errors of the (corrected) P3M energies,
  for the same test system, mesh size and real-space cut-off as in
  Fig.~1. The minimal error due to direct and reciprocal space cut-offs
  is shown as a dashed line.
  }
\end{center}
\end{figure}

The pronounced minimum in the RMS error curves stresses the importance of
using the optimal value of~$\alpha$ when performing simulations with
the P3M algorithm (or with the other variants of mesh based Ewald
sums). Our accurate RMS error estimate for the P3M energies can be
used to quickly find the optimal set of parameters (mesh size, charge
assignment order, Ewald splitting parameter) that lead to the desired
accuracy with a minimum of computational effort~\cite{DH2}. Whatever
the chosen parameters, it can serve also as a valuable indicator of
the accuracy of the P3M energies.





\section{Conclusions}

In this article, we discussed in detail which ingredients are
necessary to utilize the P3M algorithm to compute accurate Coulomb
energies of point charge distributions. The usage of a nearly linear
scaling method ($\approx N \log N$) like P3M is almost compulsory for
systems containing more than a few thousand charges.

In particular, we derived the cut-off corrections for the standard
Ewald sum transparently and interpreted the systematic errors in terms
of Madelung energies. This route lead us to an additional real-space
cut-off correction term that has so far not been discussed in the
literature.  Building on these results, we have deduced the $k$-space
cut-off correction term in the case of the P3M algorithm, where
additional aliasing errors play a role.  Furthermore we derived the
exact form of the influence function that minimizes the RMS errors in
the energies, and showed that this function is not much different from
the force-optimized influence function, which \textit{a posteriori}
justifies why in most P3M implementations the usage of the
force-optimized influence function does not lead to inaccurate
results.  Based on the energy optimized influence function we derive
an accurate RMS error estimate for the energy, and performed numerical
tests on sample configurations that demonstrate the validity of our
error estimates and the necessity to include our correction terms.  We
also demonstrated that the electrostatic energy of an individual
particle in the system can be obtained in the P3M method, but at the
expense of an additional inverse fast Fourier transform.

With the help of the newly derived error estimates we can
easily tune the desired accuracy of the P3M algorithm and find suitable
parameter combinations before running
any simulation.

The P3M algorithm can be generalized along our discussed lines
to compute other long range interactions. Of particular interest are
dipolar energies, forces and torques, and the associated error
estimates for these quantities. This will be the content of a
forthcoming publication. Our P3M generalization for the energies will
be included in a future version of the molecular simulation package
Espresso \cite{limbach06a}, that is freely available under the GNU
general public license. The website \url{http://www.espresso.mpg.de}
provides up-to-date information.

\section*{Acknowledgements}
Funds for this research were provided by the Volkswagen Stiftung under
grant I/80433 and by the DFG within grant Ho-1108/13-1.

\appendix

\section{Ewald pair potential and Madelung self-energy}	
\label{AA}

The Ewald formula for the electrostatic energy~$E$ of a periodic
charged system can be written in a form that underlines the fact that
$E$ includes the Madelung self-energies $Q^2 \zeta/2$ of the ions
[$Q^2=\sum_i q_i^2$ and $\zeta$ is defined in~\eqref{zeta}]. We recall
from Sec.~\ref{Ewald} that the Ewald formula for $E$ reads, if the
system is globally neutral and if we employ metallic boundary
conditions,
\begin{equation}
E = \frac{1}{2}\sum_{i,j}
\sideset{}{^{'}}\sum_{\vn\in\ZZ^3}q_i q_j \psi(\ver_{ij}+\vn L)
+ \frac{1}{2L^3}\sum_{i,j} q_i q_j \sum_{\substack{\vk\in\KK\\\vk\neq0}}e^{-i\vk\cdot(\ver_i-\ver_j)} \TF{\phi}(\vk)
-Q^2 \frac{\alpha}{\sqrt{\pi}}.
\end{equation}
The ``self-energy terms'' in~$E$, \ie{} term $E^{(s)}$ and terms $i=j$, are
\begin{equation}
\frac{Q^2}{2}\Big( \sum_{\vn\neq 0} \psi(\vn L) + \frac{1}{L^3} \sum_{\vk\neq0} \TF{\phi}(\vk) - \frac{2\alpha}{\sqrt{\pi}} \Big)
= \frac{Q^2}{2}\Big( \zeta + \frac{\pi}{\alpha^2 L^3} \Big).
\end{equation}
We can write therefore
\begin{align} \notag
E &= \frac{1}{2}\sum_{i\neq j} q_i q_j \Big( \sum_{\vn\neq 0} \psi(\ver_{ij}+\vn L) + \frac{1}{L^3} \sum_{\vk\neq0} e^{-i\vk\cdot\ver_{ij}} \TF{\phi}(\vk) \Big)
+ \frac{Q^2}{2}\Big( \zeta + \frac{\pi}{\alpha^2 L^3} \Big) \\	\label{def_V_Ewald}
&= \frac{1}{2}\sum_{i\neq j} q_i q_j  V_\mathrm{Ewald}(\ver_{ij}) + \frac{Q^2}{2} \zeta
\end{align}
where we defined the Ewald pair interaction~\cite{deLeeuw}
\begin{equation}
V_\mathrm{Ewald}(\ver) = \sum_{\vn\neq 0} \psi(\ver+\vn L) + \frac{1}{L^3} \sum_{\vk\neq0} e^{-i\vk\cdot\ver} \TF{\phi}(\vk) - \frac{\pi}{\alpha^2 L^3}.
\end{equation}
Notice that in writing \eqref{def_V_Ewald}, we used $\sum_i\sum_{j\neq
  i} q_i q_j\, \pi/(\alpha^2 L^3)= -Q^2\, \pi/(\alpha^2 L^3)$ which
follows from electro-neutrality. Thanks to the inclusion of this
constant in the definition of $V_\mathrm{Ewald}(\ver)$, the Ewald pair
potential does not depend on the parameter~$\alpha$
[$\partial/\partial\alpha\, V_\mathrm{Ewald}(\ver)=0$] and its average
over the simulation box is zero~\cite{Hummer98}:
\begin{equation}
\moy{V_\mathrm{Ewald}(\ver)} = \frac{1}{L^3}\int_{L^3}\dd^3\ver \, V_\mathrm{Ewald}(\ver) = 0.
\end{equation}
The latter property is simply a consequence of
$\moy{\exp(i\vk\cdot\ver_{ij})}=\delta_{\vk,\vect{0}}$ and
Eq.~\eqref{int_psi}.

In conclusion, expression~\eqref{def_V_Ewald} shows explicitly that
the electrostatic energy of a periodic charged system includes the
Madelung self-energies $Q^2 \zeta/2$ of the ions~\cite{Brush,Nijboer}.
The fact that the Ewald interaction between a pair of particles
averages to zero when one particle explores the whole simulation box
is also noteworthy aspect of Ewald potential~\cite{Hummer98}.

\section{Proof  of  Eq.~\eqref{SAMPLING_THM}}\label{AppA}
Eq.~\eqref{SAMPLING_THM} is a consequence of the Sampling theorem
[refs] and is straightforward to demonstrate. The sum
in~\eqref{FFT_rho} is rewritten as an integral
\begin{equation} 
\label{18} 
\TF{\rho}\M(\vk) =
h^3\int\dd\ver'\int_V\dd\ver \,
\cha(\ver)U(\ver-\ver')\rho(\ver')e^{-i\vk\cdot\ver} 
\end{equation} 
where we used
\eqref{charge_assignment} and introduced an infinite mesh of Dirac
delta functions 
\begin{equation} 
\label{cha} 
\cha(\ver) = \sum_{\ver_m\in\MM_\mathrm{p}}\delta(\ver-\ver_m) = 
\frac{1}{h^3}\sum_{\vm\in\ZZ^3} e^{-i k_g\vm\cdot\ver}.  
\end{equation} 
(We recall that $k_g = 2\pi/h$). Using the
above representation of $\cha(\ver)$ and introducing in~\eqref{18} the
Fourier series representation of the periodic charge density, 
\begin{equation}
\rho(\ver')=\frac{1}{L^3}\sum_{\vk'\in\KK}\TF{\rho}(\vk') \,
\exp(i\vk'\cdot\ver'), 
\end{equation} 
we recover the result~\eqref{SAMPLING_THM}
after straightforward simplifications.

\section{Proof of equivalence between Eqs. \eqref{14} and \eqref{E^K_NBI}}\label{AppB}

Eq.~\eqref{14} is equivalent to 
\begin{equation} 
\label{A.14}
E^{(ks)}_{\PPPM}=\frac{1}{2V}\sum_{\substack{\vk\in\KK\\
    \vk\neq0}}\TF{\rho}^*(\vk) \TF{\Phi}(\vk), 
\end{equation} 
where
$\TF{\Phi}(\vk)$ is the \emph{full} Fourier transform ($\vk\in\KK$) of
the back-interpolated potential mesh~\eqref{Phi(r)}: 
\begin{equation}
\TF{\Phi}(\vk) = \int_V \Phi(\ver) \dd\ver = h^3\int_V\dd\ver \,
e^{-i\vk\cdot\ver} \int\dd\ver' \cha(\ver') U(\ver-\ver')
\Phi\M(\ver').  
\end{equation} 
We replace in this equation $\Phi\M(\ver')$ and
$\cha(\ver')$ by their expressions \eqref{Phi_M(r)} and \eqref{cha},
and perform the integration over~$\ver'$: 
\begin{equation} 
\TF{\Phi}(\vk) =
\frac{1}{L^3}\sum_{\vk'\in\TFM} \TF{\Phi}\M(\vk')
\sum_{\vm}\int_V\dd\ver \, e^{-i\vk\cdot\ver} \, \TF{U}(\vk'+k_g\vm)
e^{i(\vk'+k_g\vm)\cdot\ver}.  
\end{equation} 
The integration over~$\ver$
introduces a Kronecker symbol $\delta_{\vk,\vk'+k_g\vm}$. We get
therefore the simple result 
\begin{equation} 
\label{TF_Phi} 
\TF{\Phi}(\vk) =
\TF{\Phi}\M(\vk) \TF{U}(\vk)
\end{equation} 
where the function $\TF{\Phi}\M(\vk)$, which is defined originally
only for $\vk\in\TFM$, is now understood to be extended periodically
to all $\KK$ space. Notice that the inverse FFT does not introduce
aliasing errors: the sum over $\vm$ merely renders $\TF{\Phi}\M(\vk)$
periodic. In accordance with \eqref{FFT_rho} and \eqref{def_G}, we
extend also $\TF{\rho}\M(\vk)$ and $\TF{G}(\vk)$ periodically, with
period $2\pi/h$.  Using the above result and~\eqref{def_G}, the
reciprocal energy~\eqref{A.14} can be expressed as
\begin{align}
   E^{(ks)}_{\PPPM} &=\frac{1}{2L^3}\sum_{\vk\in\KK} \TF{\rho}^*(\vk)\TF{U}(\vk)
   	\TF{\rho}\M(\vk)  \TF{G}(\vk)\\
&= \frac{1}{2L^3}\sum_{\vk\in\TFM}\sum_{\vm\in\ZZ^3} \TF{\rho}^*(\vk+k_g\vm)\TF{U}(\vk+k_g\vm)
   	\TF{\rho}\M(\vk)  \TF{G}(\vk).
\end{align}
This may be compared with Eq.~\eqref{E^K_NBI}, \ie{}
\begin{equation} 
\label{E^k_bi}
E^{(ks)}_{\PPPM}=\frac{1}{2L^3}\sum_{\substack{\vk\in\TFM\\
    \vk\neq0}}\TF{\rho}\M^*(\vk) \TF{\rho}\M(\vk)\TF{G}(\vk).  
\end{equation}
Recalling~\eqref{SAMPLING_THM} and the fact that $\TF{U}(\vk)$ is
real, we see that both expressions are equivalent.


\end{document}